\documentclass[aps,prb,twocolumn,groupedaddress,showpacs]{revtex4}
\usepackage{graphicx}
\usepackage{amsmath,amsthm}

\begin{document}


\title{
Structural properties  in Sr$_{0.61}$Ba$_{0.39}$Nb$_{2}$O$_{6}$  in the temperature range 10 K to 500 K investigated by high-resolution neutron powder diffraction and specific heat measurements}

\author{J. Schefer$^1$, D. Schaniel$^2$, V. Pomjakushin$^1$, U. Stuhr$^1$, V. Pet\v{r}\'{\i}\v{c}ek$^3$, Th. Woike$^2$, M. W\"ohlecke$^4$ and  M. Imlau$^4$}
\email{jurg.schefer@psi.ch}
\affiliation{$^1$Laboratory for Neutron Scattering, ETHZ \& PSI, CH-5232 Villigen, PSI, Switzerland\\
$^2$Institut f\"ur Mineralogie, University at Cologne, D-50674 K\"oln, Germany\\
$^3$Institute of Physics, Academy of Sciences of the Czech Republic, Na Slovance 2,18221 Praha 8, Czech Republic\\
$^4$Fachbereich Physik, University of Osnabr\"uck, D-49069 Osnabr\"uck, Germany}

\date{\today}

\begin{abstract}

We report high-resolution neutron powder diffraction on Sr$_{0.61}$Ba$_{0.39}$Nb$_{2}$O$_{6}$ , SBN61, in the temperature range 15-500\,K. The results indicate that the low-temperature anomalies (T $\leq$ 100\,K)  observed in the dielectric dispersion are due to small changes in the incommensurate modulation of the NbO$_6$ octahedra, as no structural phase transition of the average structure was observed. This interpretation is supported by specific heat measurements, which show no latent heat, but a glass-like behavior at low temperatures. Furthermore we find that the structural changes connected with the ferroelectric phase transition at $T_c \simeq 350$\,K start already at 200\,K, explaining the anisotropic thermal expansion in the temperature range 200-300\,K observed in a recent x-ray diffraction study.
\end{abstract}

\pacs{61.12.Ld 33.15.Dj}

\maketitle

\section{Introduction}

Strontium barium niobate (Sr$_{x}$Ba$_{1-x}$Nb$_2$O$_{6}$, SBNx) exhibits very large electro-optic, piezoelectric, and pyroelectric coefficients\cite{Glass-1969, Ewbanks-1987} and is therefore of significant interest for a variety of applications, especially as a photorefractive material with a high two-beam coupling coefficient $\Gamma$ \cite{Ewbanks-1987,woike-APB-2001}. Furthermore, SBN is a model substance for the investigation of the relaxor type  ferroelectric phase transition, where ferroelectric nanoclusters are stabilized by  internal random fields above the critical temperature $T_c$ over a wide temperature range, such that the ferroelectric polarization does not decay spontaneously at $T_c$ \cite{Bhalla-PRB-1987}. This relaxor behavior is well explained by the Random-Field-Ising model for the ferroelectric phase transition. Assuming an internal random electric field, all critical exponents could be determined according to the scaling relation \cite{Dec-EL-2001,Granzow-PRL-2004, Kleemann-JMS-2006}. They fulfill the Rushbroke relation and belong to the universal class of the three-dimensional  Random-Field-Ising model. During this high-temperature ferroelectric phase transition the space group of the average structure changes from $P-4b2$ to $P4bm$, while the incommensurate modulation remains \cite{schneck1981,Balagurov-sbn-1987}. 
\begin{figure} \includegraphics[width=7cm,angle=0]{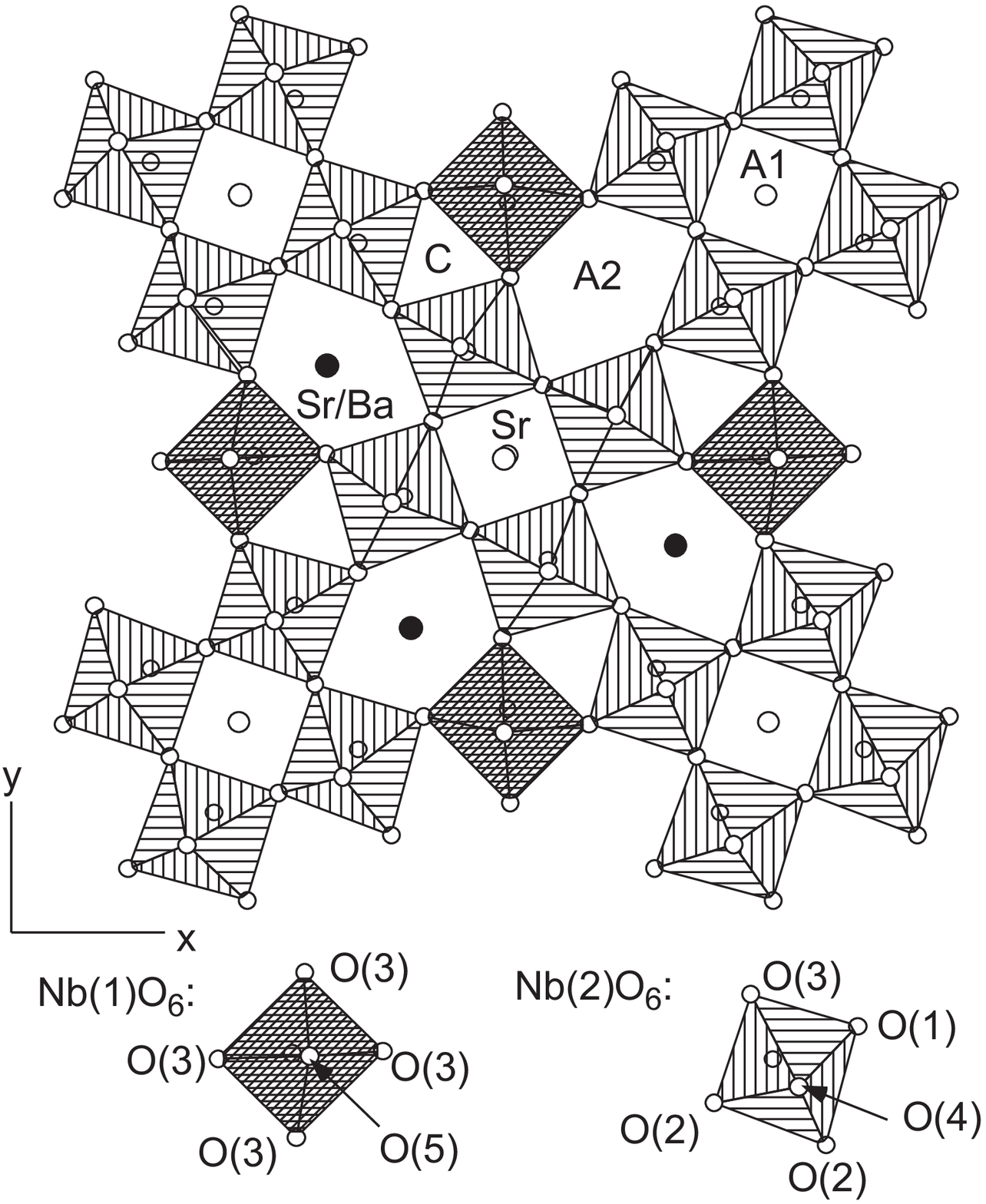}%
\caption{Projection of Sr$_{0.61}$Ba$_{0.39}$Nb$_{2}$O$_{6}$ along the c-axis. The pentagonal channels A2 are filled by Sr/Ba, the tetragonal channels A1 by Sr, and the trigonal channels C remain empty. 5 Sr/Ba atoms are distributed over 6 A1/A2 sites.}
\label{SBN61-struktur}
\end{figure}

SBN belongs to the large class of materials with open tungsten bronze structure. This structure is characterized by a network of NbO$_6$ octahedra, which are connected with  the edges to form pentagonal A2, tetragonal A1, and trigonal C channels (see Fig.\,\ref{SBN61-struktur}). The 12-fold coordinated A1 position is occupied only by Sr, while the 15-fold coordinated A2 site is shared between Sr and Ba. The C positions remain empty. Since only 5 Sr/Ba atoms are occupying the 6 A1/A2 sites, one of the A1/A2 sites remains empty. The incommensurate structure \cite{Woike-SBN61-2003} is due to a tilting of the oxygen octahedra with respect to the c-axis. Neglecting the incommensurate nature of the modulation this bending was earlier described by two configurations for the oxygen octahedra \cite{Chernaya-Barium-1997}.

There is an ongoing discussion about a second phase transition at  temperatures below 100 K  \cite{Xu-PRB-1989, prokert-SBN39-1995, Ko-JAP-2002, Faria-JRS-2003, buixaderas-sbn-modulated-structure-2005}. Based on dielectric, pyroelectric, and structure measurements, a phase transition from the tetragonal (4mm) phase into an orthorhombic or monoclinic (m) phase was postulated \cite{Xu-PRB-1989}. This postulation is supported by temperature dependent dielectric and pyroelectric signals in [100] and [110] directions, which are forbidden in the tetragonal 4mm point group. On the other hand in an early x-ray powder diffraction study \cite{prokert-SBN39-1995} only small changes in the lattice parameters a and c were detected without indications for a phase transition into a monoclinic phase.  Later, Ko \textit{et al.} \cite{Ko-JAP-2002} confirmed the dielectric measurements, but argued, that the signals are not necessarily due to a structural phase transition, but  might as well be due to a temperature driven charge or polaronic hopping process, or even due to a temperature driven flipping process of the oxygen octahedra, which should be reflected in the incommensurate positions of the oxygen atoms. The activation energy for this process was given as 0.087 eV. More recent evidence for a structural phase transition was claimed from Raman spectroscopy \cite{Faria-JRS-2003}, while a following IR study gave no clear evidence for a new phase transition below 100\,K \cite{buixaderas-sbn-modulated-structure-2005}.

In order to clarify whether the observed dielectric anomalies below 100 K are due to a structural phase transition, we performed temperature dependent neutron diffraction experiments on powders of SBN61, and specific heat measurements on single crystals of SBN61. Neutron diffraction offers the advantage of a high sensitivity to oxygen and hence allows for a detailed investigation of the influence of the NbO$_6$ octahedra on the observed physical properties. The neutron diffraction experiments were carried out in the temperature range 15-500 K, while the specific heat was recorded from 2 to 300 K. We could not observe signatures of a structural phase transition, which supports the interpretation of Ko \textit{et al.} \cite{Ko-JAP-2002} that a dynamical process, like the concerted rotation of the oxygen octahedra, is at the origin of the observed phenomena.

\section{Experimental Details}

The congruent SBN61 crystals were grown by the Czochralski method in the crystal growth laboratory of
the University of Osnabr\"uck. The polycrystalline sample was produced by grinding a  single crystal.

Specific heat measurements were performed in the temperature range 2-300\,K by the relaxation method using a PPMS from Quantum Design. Good thermal contact between the single crystal sample and the sapphire chip of the system was made with apiezon grease. 

The neutron powder diffraction experiments were performed on the high-resolution powder diffractometer HRPT \cite {fischer-hrpt-2000} at the Swiss Spallation Neutron Source SINQ
\cite {fischer-w-e-sinq-1997}, using $\lambda = 1.886$\,\AA. Temperature dependent measurements were done in the high intensity mode. Additionally, two patterns at T = 15\,K and T = 290\,K have been recorded using the high-resolution mode (primary collimation $\alpha_1$=12\') on HRPT. In order to determine the lattice constants with high accuracy further powder diffraction experiments were performed on the  high resolution neutron strain scanner POLDI \cite{Stuhr-2005-POLDI} in the temperature range 300-500\,K. A chopper running at 10'000 rpm allows for the analysis of the frame overlapping time-of-flight data.  Apart from the (620) and (002) reflections, we only analysed fully separated reflections in order to get the highest accuracy. The concept of the instrument is described in detail by Stuhr \cite{Stuhr-2005-POLDI-concept}.

\section{Results}

\subsection{Specific heat}

\begin{figure} \includegraphics[width=8cm,angle=0]{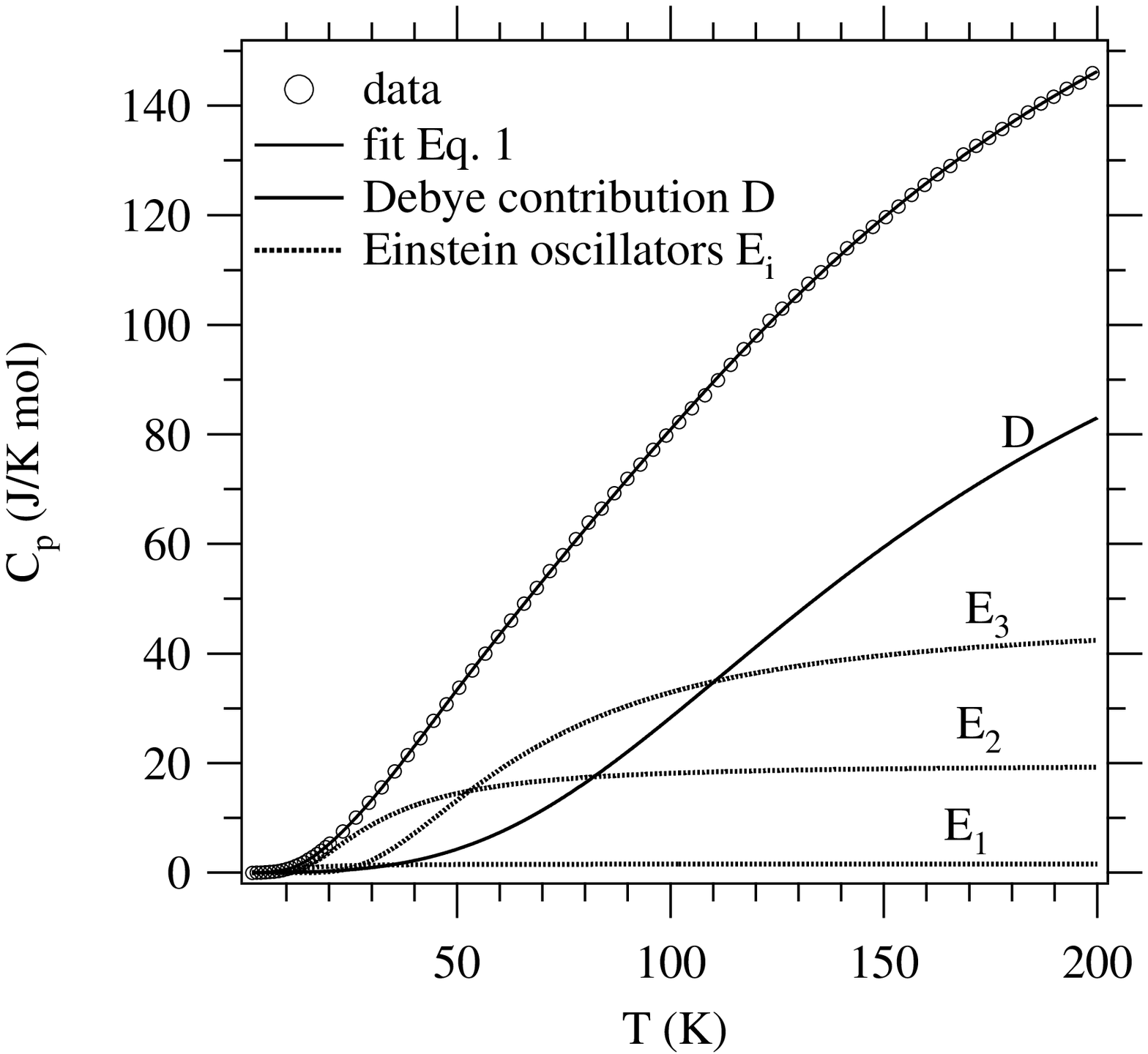}%
\caption{Specific heat in the temperature range 2-200\,K for Sr$_{0.61}$Ba$_{0.39}$Nb$_{2}$O$_{6}$. The solid line is a fit using Eq.\,\ref{eq:cp}, the dashed and dash-dotted lines correspond to the Einstein and Debye contributions, respectively.}
\label{SBN61-cp}
\end{figure}

\begin{figure} \includegraphics[width=8cm,angle=0]{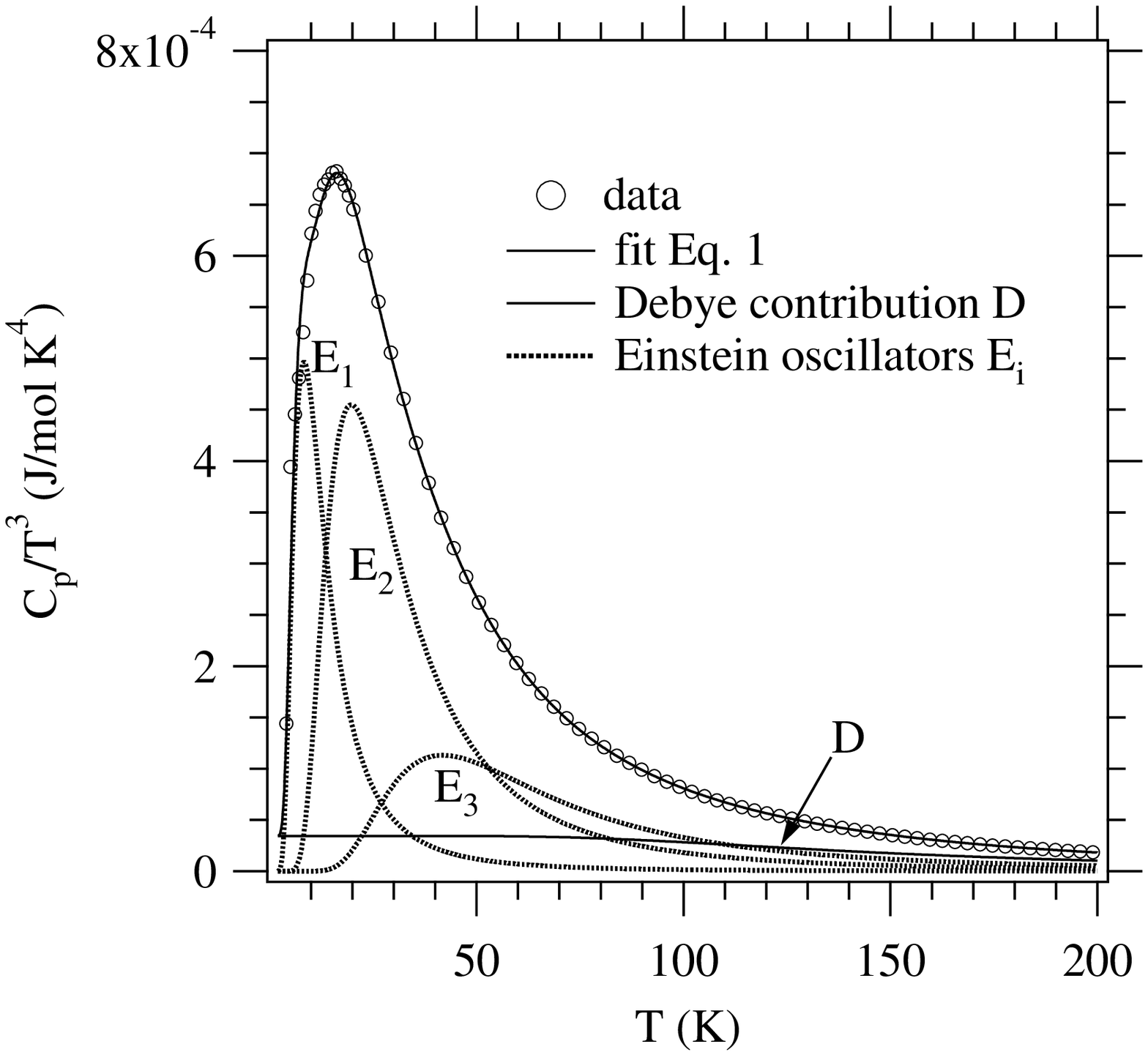}%
\caption{Specific heat $C_{p}/T^3$ in the temperature range 2-200\,K for Sr$_{0.61}$Ba$_{0.39}$Nb$_{2}$O$_{6}$. The solid line is a fit using Eq.\,\ref{eq:cp}, the dashed and dash-dotted lines correspond to the Einstein and Debye contributions, respectively.}
\label{SBN61-cp2}
\end{figure}

The specific heat of SBN61 was measured in the temperature range 2-300\,K (Fig.\,\ref{SBN61-cp}). No latent heat was observed in this temperature range. The data were analyzed using the well known Debye relation and three Einstein oscillators, such that 
\begin{equation}
\label{eq:cp}
\begin{split}
C_{p} = & 9 n_D N_{A}k_{B}\left(\frac{T}{\theta_D}\right)^3\int_0^{x_D}dx\frac{x^4e^x}{(e^x-1)^2} +\\ 
& + 3N_{A}k_{B}\sum_{i=1}^3 n({E_i})\left(\frac{E_i}{T}\right)^2\frac{e^{E_i/T}}{(e^{E_i/T}-1)^2},
\end{split}
\end{equation}
where $x_D = \theta_D/T = \hbar\omega_D/k_BT$ with $\theta_D$ the Debye-temperature, $k_B$ the Boltzmann constant, $N_A$ the Avogadro number, and $\omega_D$ the Debye frequency. The parameter $n_D$ denotes the relative strength of the Debye contribution, $n(E_i)$ that of the single Einstein oscillators, and $E_i$ is the energy of the Einstein oscillators. Three Einstein oscillators were necessary since a fit with only two was not accurate in the low temperature part. The fit is performed up to 200 K, since the structural data show, that above this temperature the relaxor phase transition sets in (see below). From Figure \ref{SBN61-cp} it becomes clear, that the Debye contribution is dominant above about 110 K, below the Einstein oscillators are mainly contributing to the total specific heat. The contributions of the three Einstein oscillators is more clearly visible in Fig.\,\ref{SBN61-cp2}, where the specific heat is plotted as $C_{p}/T^3$ versus $T$. The fit parameters are given in Table \ref{cpfit}. 

\begin{table}
  \centering 
  \caption{Parameters of Einstein oscillators and Debye contribution obtained from a fit of Eq.\,\ref{eq:cp} to the specific heat data in the temperature range 2-200\,K.}\label{cpfit}
 {\smallskip}
  \begin{tabular}{cccc}
\hline
	$i$  & $E_i$ (K) & $\nu_i$ (cm$^{-1}$) & $n(E_i)$ \\
        \noalign{\smallskip}\hline\noalign{\smallskip}
      1 & 41(19) & 29(12) & 0.06 (8)   \\
      2 & 97(13) & 67(10) & 0.78(12)   \\
      3 & 206(11) & 143(9) & 1.85(12)  \\
        \noalign{\smallskip}\hline\noalign{\smallskip}
	 & $\theta_D$ (K) & $\omega_D$ (cm$^{-1}$) & $n_D$  \\
        \noalign{\smallskip}\hline\noalign{\smallskip}
    & 681(8) & 474(6) & 5.6(1)  \\
   \noalign{\smallskip}\hline
 \end{tabular}
\end{table}

\subsection{Neutron diffraction: Temperature dependence}

\begin{figure} \includegraphics[width=8cm]{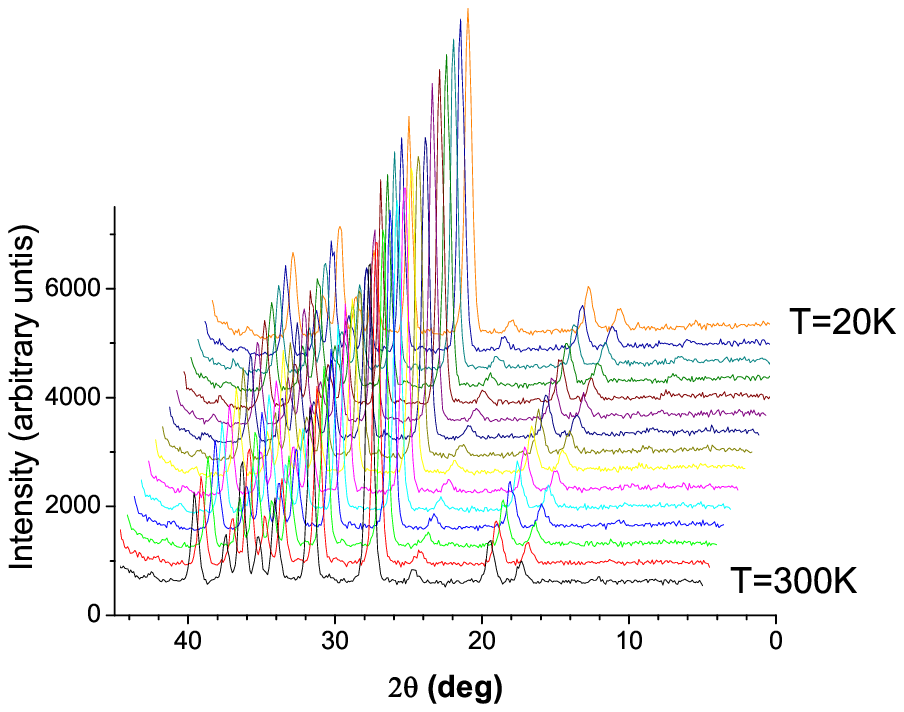}%
\caption{(Color online) Section of the neutron diffraction patterns of SBN61 collected on HRPT using the high-intensity mode, $\lambda = 1.886$\,\AA,for T=300K down to T=20K in steps of 20K.}
\label{SBN61-pattern-temperature}
\end{figure}

\begin{figure} \includegraphics[width=8cm]{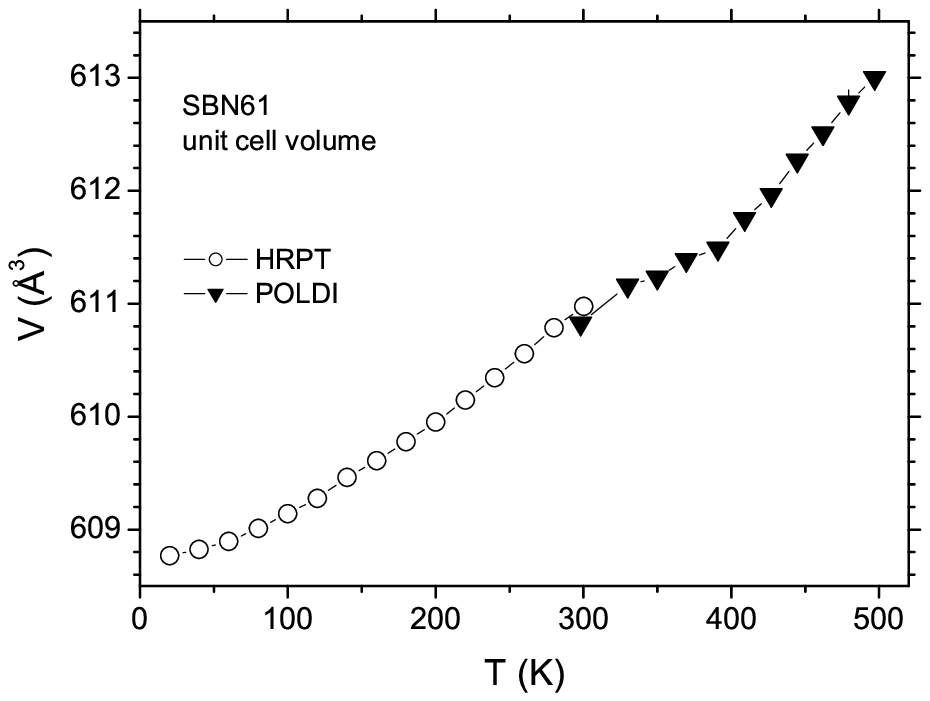}%
\caption{Volume of the tetragonal unit cell in SBN61 as a function of temperature. Open circles correspond to HRPT data, closed triangles
to POLDI data.}
\label{SBN61-volume}
\end{figure}

\begin{figure} \includegraphics[width=8cm]{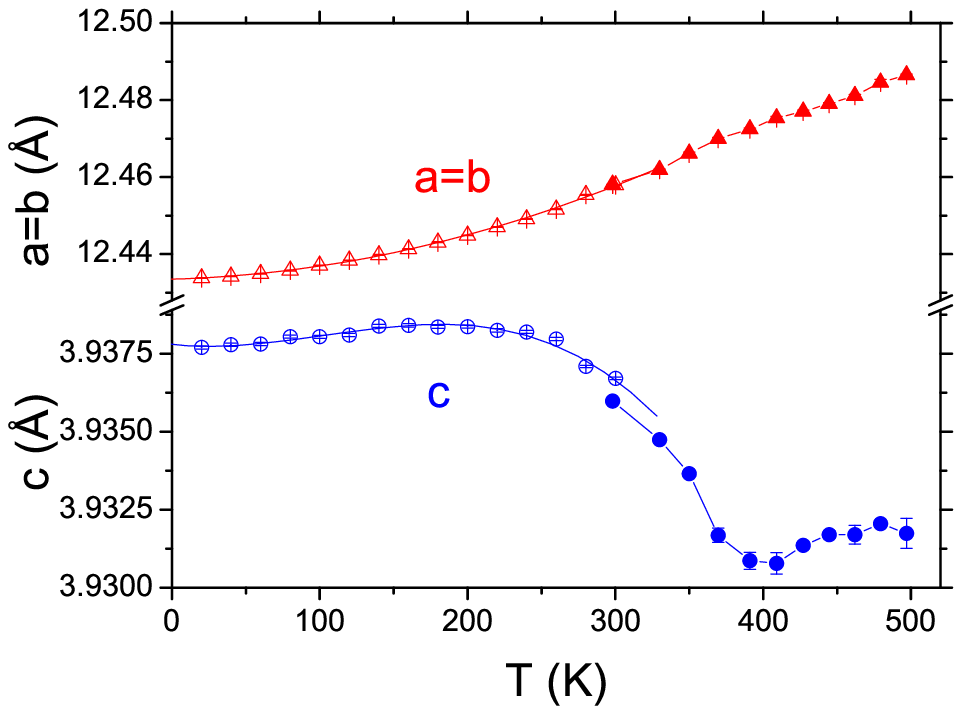}%
\caption{Tetragonal lattice constants a and c of SBN61 as a function of temperature. Open circles correspond to HRPT data (T $\leq$ 300K), closed circles
to POLDI data(T $\geq$ 300K).}
\label{SBN61-abc}
\end{figure}

\begin{figure} \includegraphics[width=8cm,angle=0]{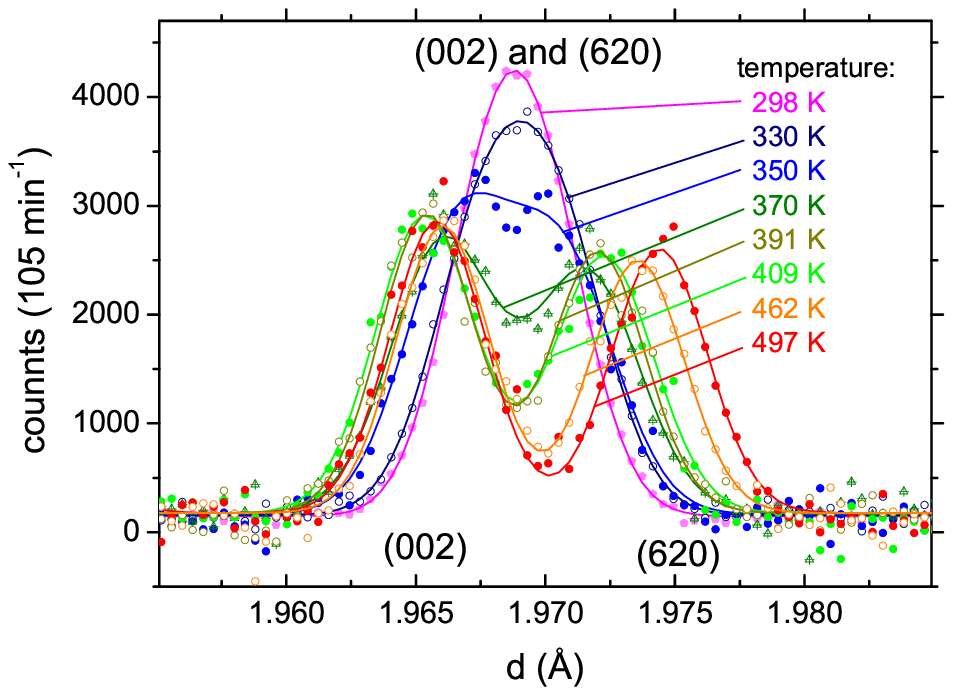}%
\caption{(Color online) Splitting of the (620) and (002)reflections in  Sr$_{0.61}$Ba$_{0.39}$Nb$_{2}$O$_{6}$, SBN61, observed with the high resolution
strain scanner POLDI \cite {Stuhr-2005-POLDI}.  At low temperatures, the two reflections nearly coincide, but split a higher temperatures 
as a result of the opposite temperature dependence of a=b and c.}
\label{SBN61-splitting-POLDI}
\end{figure}

In order to detect a possible low-temperature structural phase transition we  collected neutron diffraction powder patterns on HRPT between 10 K and 300K in the high intensity mode. Fig.\,\ref{SBN61-pattern-temperature} shows the low angle part of these patterns, indicating that no structural phase transition takes place. In addition we collected two data sets in the high resolution mode at 15\,K and 290\,K. Also here no indication for a structural change was observed. Lattice parameters were determined from these powder patterns using the average structure $P4bm$. The volume of the tetragonal unit cell is steadily increasing below 300\,K, as illustrated in Fig.\,\ref{SBN61-volume}. Between 300\,K and 400\,K the high temperature ferroelectric phase transition is clearly seen from the change in the lattice constant c as illustrated in \ref{SBN61-abc}. Fitting a polynom $V(T) = V_0+\alpha_VT+\beta_VT^2$ to the measured data below 300\,K yields the parameters $\alpha_V = 3.7(3)\cdot10^{-3}$\,\AA$^{3}$K$^{-1}$, $\beta_V = 1.37(9)\cdot10^{-5}$\,\AA$^{3}$K$^{-2}$, and the volume $V_0 = V(T = 0) = 608.65(2)$\,\AA$^{3}$.  The lattice constant \textit{a} \, follows the same tendency, whereas the tetragonal \textit{c}-axis is decreasing for T$>$200K (see Fig. \ref{SBN61-abc}). In order to follow the change of the lattice parameters through the ferroelectric phase transition ($T_c \simeq 350$\,K) we performed further experiments on POLDI in the temperature range 300-500\,K. Here the lattice parameters are obtained with high accuracy from separated reflections, as illustrated in Fig.\,\ref{SBN61-splitting-POLDI} for the (620) and (002) reflections. In the ferroelectric $P4bm$ phase the two reflections accidentally overlap. As a result of the different temperature dependences of the a and c parameter in the high-temperature range they are splitted. Note that in Figs.\,\ref{SBN61-volume} and \ref{SBN61-abc} the lattice parameters obtained from the two different instruments were not corrected for the calibration 
difference of 0.1$\%$. 
In total the lattice constant \textit{a} shows the usual temperature dependence as it increases steadily below 300\,K and changes its slope above 300\,K. A polynomial fit $a(T) = a_0+\alpha_aT+\beta_aT^2$ up to 300\,K yields the parameters $\alpha_a = 1.0(4)\cdot10^{-5}$\,\AA K$^{-1}$, $\beta_a = 2.4(1)\cdot10^{-7}$\,\AA K$^{-2}$, and $a_0 =  12.4340(2)$\,\AA. The lattice constant \textit{c} on the other hand increases in a similar manner up to about 150\,K, where a flattening sets in. Above about 200\,K up to 400\,K \textit{c} decreases, and above 400\,K it increases again. There is an indication for a second decrease above 500\,K, but further measurements at higher temperatures are necessary to confirm this observation, which also has been observed in SBN82 \cite{Podlozhenov-2006}.

\subsection{Neutron diffraction: Structural Refinement}

\begin{figure} \includegraphics[width=9cm,angle=-0]{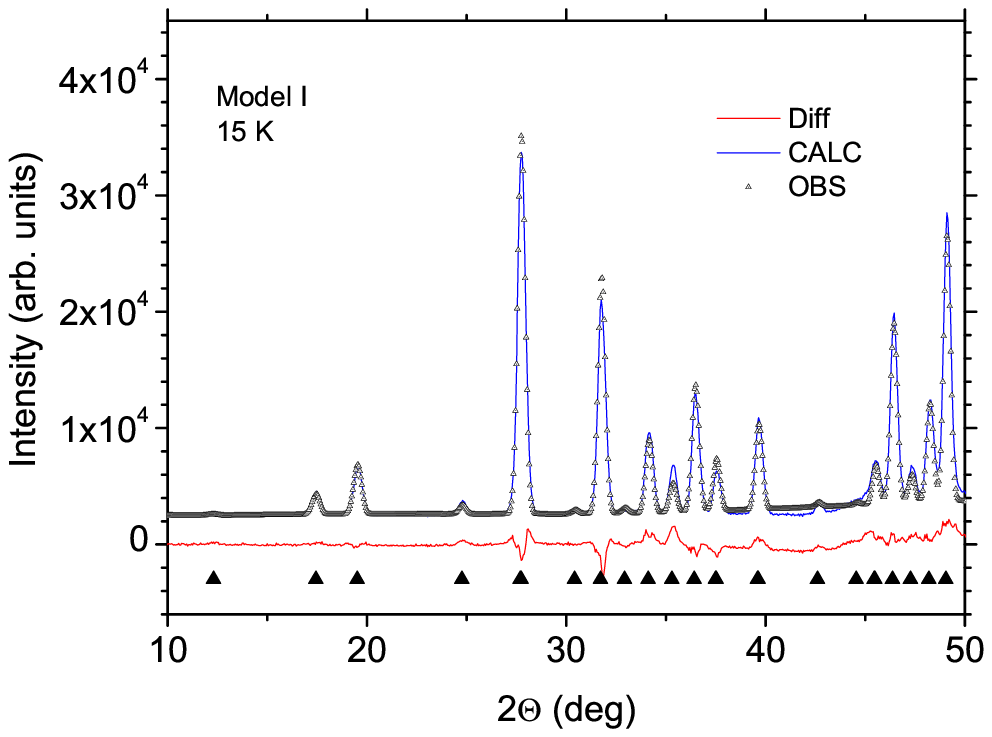}%
\caption{(Color online) Section of the graphical result of the Rietveld \cite {petricek2000} refinement of the average structure (model I) of SBN61 powder data collected on HRPT using the high-resolution mode, $\lambda = 1.886$\,\AA, at T=15\,K (column 1, table \ref {SBN61-summary-15K-290K}).}
\label{SBN61-jana2000-15K-averaged}
\end{figure}

\begin{table*}
\caption{Comparison of different refinements for  Sr$_{0.61}$Ba$_{0.39}$Nb$_{2}$O$_{6}$, SBN61, measured on HRPT/SINQ at $\lambda = 1.886$\,\AA \, in the high-resolution mode. Tetragonal space group P4bm/X4bm \cite{Yamamoto-1996-quasiperiodic}(No.100), Z=8, \textbf{q}$_{1,2}$ = (.3075,$\pm.3075$,0)\cite{Woike-SBN61-2003}. Atoms are located on sites 2b(Nb1), 8d (Nb2),2a (Sr1),4c (Sr2,Ba@Sr2), 4c (O1),8d (O2,O3,O4), 2b (O5). The \textit{c}-axis of the incommensurate 5-dimensional
space group X4bm is doubled compared to the P4bm in order to account for modulation along \textit{c} with q$_z$=0.5. Structure refinement  was using JANA2000   \cite {petricek2000}. Detailed results of the refinements using a positional modulated structure are given in Tabs. V-XII in the supplementary material \cite{Schefer-Supplementary-Material-2006}.}
\label{SBN61-summary-15K-290K}
{\smallskip}
\begin{tabular}{l|lll|lll}
\hline
	$T$ (K)                & 15        & 15           & 15              & 290           & 290          & 290                    \\
Model		    & I  & II    & III       & I     & II    & III              \\
Positional modulation    & averaged  & all sites    & all sites       & averaged      & all sites    & all sites              \\
ADP modulation           &      -    & No           & site Sr2/Ba&      -        & No           &    site Sr2/Ba    \\
\hline
	 a (\AA)               & 12.4387(3)& 12.4383(2)   &   12.4383(2)    & 12.4626(3)    & 12.4619(2)   & 12.4620(2)             \\
c$'$=2 $\cdot$ c (\AA)        &  7.8789(3)&  7.8787(2)   &    7.8788(2)    &  7.8773(2)    & 7.8772(2)    &  7.8772(2)             \\
2 $\cdot$Volume (\AA$^3$)& 1219.05(6)&  1218.96(4)  & 1218.96(4)      & 1223.48(5)    & 1223.31(4)   &  1223.34(4)            \\
\hline
 $ \chi^2   $              & 51.11     &\textbf{22.27}& 21.76      & 24.84         &\textbf{12.54}& 12.19                  \\
	\hline
	main reflections  & & & & & & \\
	\hline 
	R$_{obs}$       &  7.34     &  2.91        & 2.83            & 6.62          & 2.93         & 2.88                   \\
	R$_{w,obs}$   &  5.42     &  2.61        & 2.57            & 4.99          & 2.66         & 2.61                   \\
	R$_{all}$         &  7.28     &  2.91        & 2.83            & 6.69          & 2.93         & 2.88                   \\
	R$_{w,all}$     &  5.42     &  2.61        & 2.57            & 4.99          & 2.66         & 2.61                   \\
	n$_{obs}$       &  237      &  238         & 238             & 236           & 238          & 238                    \\
	n$_{all}$         &  238      &  238         & 238             & 238           & 238          & 238                    \\
	\hline
	satellites of order 1  & & & & & & \\
	\hline 
  R$_{obs}$                &     -     & 4.53         & 4.49            & -             & 4.82        & 4.64                   \\
	R$_{w,obs}$     &     -     & 3.02         & 2.95            & -             & 3.00        & 2.87                   \\
	R$_{all}$           &     -     &  4.89        & 4.81            & -             & 5.03        & 4.95                   \\
	R$_{w,all}$       &     -     & 3.02         & 2.95            & -             & 3.00        & 2.87                   \\
	n$_{obs}$          &     -     & 765          & 767             & -             & 768         & 768                    \\
	n$_{all}$            &     -     & 777          & 777             & -             & 779         & 779                    \\
\hline
\end{tabular}
\end{table*}

There is no direct evidence for a low-temperature structural phase transition from the tetragonal into an orthorhombic or monoclinic phase from the above presented data. 
In order to detect more subtle structural changes we investigated the two high-resolution powder patterns measured at 15\,K and 290\,K in more detail. For this purpose we have to take into account, that SBN61 has an incommensurately modulated structure, as was determined by single-crystal x-ray \cite{Woike-SBN61-2003}and neutron \cite{Schaniel-Schefer-SBN61-superspace-2002,schaniel-thesis-2003} diffraction. However, the satellite peaks are too weak and too broad and can hence not be separated in the powder patterns. Therefore we performed the refinement in three steps. First the refinement was performed in the average structure, i.e. space group $P4bm$ without any modulation (model I). Second we introduced a positional modulation for all atoms (model II), i.e. two harmonic waves were included to describe the positional modulation of each atom according to the two modulation vectors  q$_1$ and  q$_2$. In the third step (model III) model II was extended with an additional modulation of the anisotropic displacement parameters (ADP) on sites Sr2 and Ba.
The refinement of the modulated structure has been done in 5-dimensional space in space group P4bm(pp$1/2$, p-p $1/2$) \cite{Wolff-Superspace-1981}. This elegant way to
handle incommensurate structures by higher-dimensional space is integrated in  JANA2000 \cite {petricek2000}. 
It is based on an extension of the standard formalism for reciprocal space to higher dimensions, including the incommensurate vectors (in our case: 
 \textbf{q}$_{1}$ and  \textbf{q}$_{2}$ yielding a 5-dimensional space):

\begin{equation}
\label{eq:def-rec}
\begin{split}
\mathbf{A}_{i} \cdot \mathbf{A}_{j}^*   =\delta_{ij} \\
 i,j=1-3,4,5
\end{split}
\end{equation}
where

\begin{equation}
\label{eq:def-e1}
\begin{split}
\mathbf{A}_4^* = \mathbf{e}_1 + \mathbf{q}_1
\end{split}
\end{equation}
and
 \begin{equation}
\label{eq:def-e2}
\begin{split}
\mathbf{A}_5^* = \mathbf{e}_2 + \mathbf{q}_{2} 
\end{split}
\end{equation}
 with $\textbf{e}_{1}$ and $\textbf{e}_{2}$ perpendicular to real (reciprocal) space $ \Re$. \\

\begin{figure} \includegraphics[width=9cm,angle=-0]{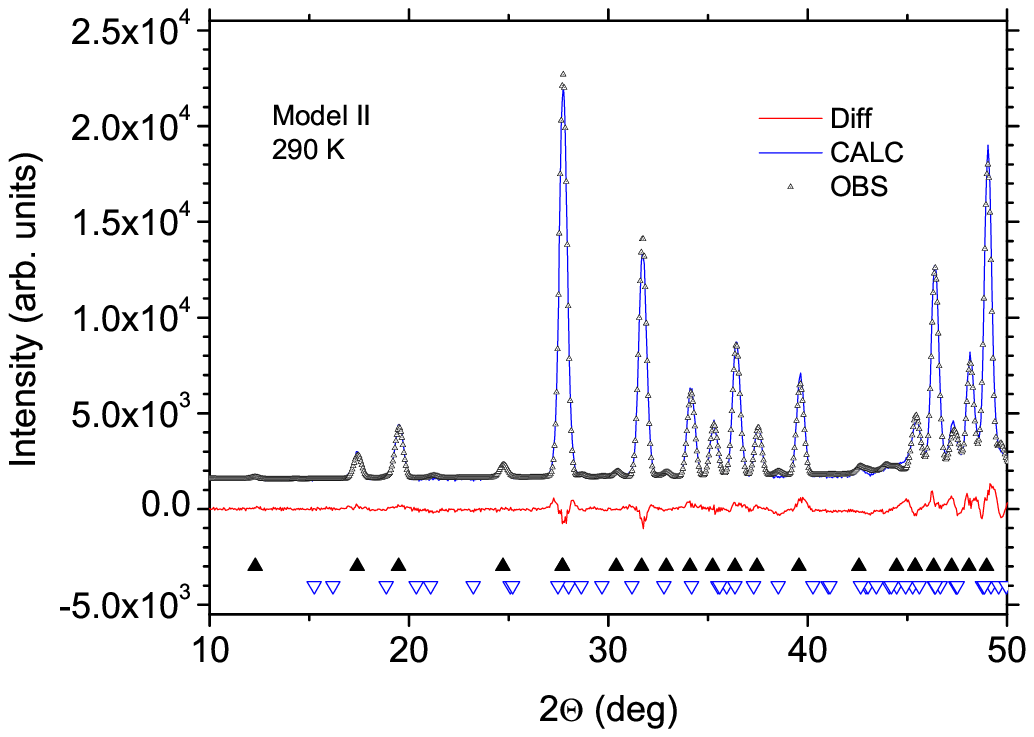}%
\caption{(Color online) Section of the graphical result of the Rietveld \cite {petricek2000} refinement of SBN61 powder data collected on HRPT using the high-resolution mode, $\lambda = 1.886$\,\AA, at temperature T=290K.  Modulations \textbf{$q_1$} and \textbf{$q_2$} have been fixed to (0.3075,$\pm0.3075$,0) using a doubled c-axis corresponding to (0.3075,$\pm0.3075$,0.5) in the original cell. 
HKL's are marked by closed triangles(average structure: top row) and by open triangles (incommensurate satellites, lower row).
Detailed refinement parameters are listed in \cite{Schefer-Supplementary-Material-2006}, Tables VII-VIII.}
\label{SBN61-jana2000-290K}
\end{figure}

\begin{figure} \includegraphics[width=9cm,angle=-0]{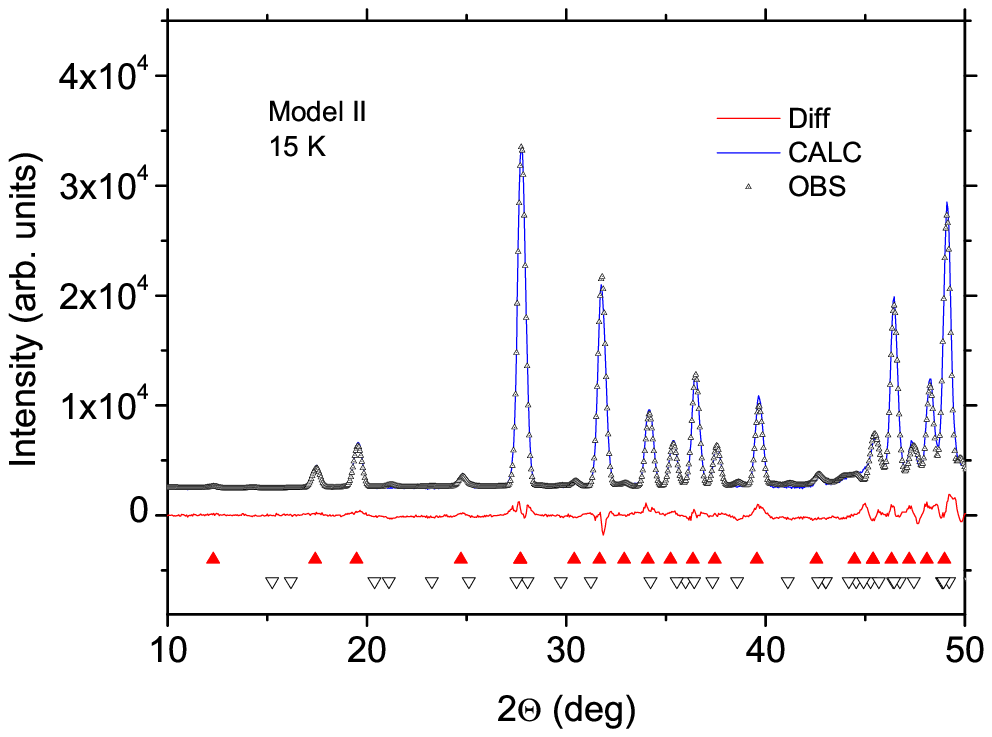}%
\caption{(Color online) Section of the graphical result of the Rietveld \cite {petricek2000} refinement of SBN61 powder data collected on HRPT using the high-resolution mode, $\lambda = 1.886$\,\AA, at temperature T=15K. Modulations \textbf{$q_1$} and \textbf{$q_2$} have been fixed to 0.3075,$\pm0.3075$,0  using a doubled c-axis. HKL's are marked by closed triangles(average structure: top row) and by open triangles (incommensurate satellites, lower row).
 The refinement  clearly improves compared to the averaged structure refinement shown in Tab. \ref{SBN61-summary-15K-290K}, model II. Detailed refinement parameters are listed in \cite{Schefer-Supplementary-Material-2006}, Tables V-VI.}
\label{SBN61-jana2000-15K}
\end{figure}

The modulation vectors \textbf{q}$_1$ and \textbf{q}$_2$  have been fixed to (0.3075,$\pm0.3075$,0) as determined from x-ray diffraction \cite{Woike-SBN61-2003}.
 A refinement of the \textbf{q}-directions is not possible as the satellites are too weak and too broad in the powder pattern. The occupation of the Ba site has been fixed to the known composition of 0.39 in all models and the sum of Ba and Sr atoms is fixed to 1. The results of the Rietveld refinements are summarized in Tab. \ref{SBN61-summary-15K-290K} and given in detail in Tabs. V-VIII and Figs. 12 and 13 in  \cite{Schefer-Supplementary-Material-2006}. The quality of the refinements are illustrated in Figs.\,\ref{SBN61-jana2000-15K-averaged}, \ref {SBN61-jana2000-290K}, and \ref{SBN61-jana2000-15K}.

Fig.\,\ref{SBN61-jana2000-15K-averaged} shows the refined pattern at 15\,K after final refinement with model I, i.e. the average structure. There are no additional peaks or splittings indicating another space group or phase. However, the refinement is not satisfactory, as peak intensities do not match very good. This is reflected also in the relatively high final  $ \chi^2$ of 51.11 and 24.84 
for the low- and high-temperature data, respectively (see Table\,\ref{SBN61-summary-15K-290K}, column 1 and 4)
and the overestimated background in the region 2$\theta$  $\sim$ 42$^0$ in Fig. \ref{SBN61-jana2000-15K-averaged}.
The fit quality is significantly better for the high temperature data with $\chi^2 = 24.84$  compared to   $\chi^2 = 51.11$  for the low-temperature data. The only significant differences in the refined parameters for the two data sets are observed in the $x$ and $y$ coordinates of the O4 atom, which change from (0.0814(3), 0.2031(3),0.221(2)) at 290\,K to (0.0830(4), 0.2017(3),0.217(2)) at 15\,K. Furthermore we note that the thermal displacement parameters of the O(4) and O(5) do not decrease with decreasing temperature and that the Sr(1) occupation is 0.65(1) and the Sr(2) occupation is 0.44(1). The parameters of the structural refinement of the average structure at 15\,K and 290\,K are listed in Table\,\ref{SBN61-compare-average}.

Introducing positional modulation waves for all atoms significantly improves the refinement as illustrated in Figs.\,\ref{SBN61-jana2000-290K} and \ref{SBN61-jana2000-15K}. The R-values decrease to $R_{obs} = 2.91$ and $R_{obs} = 2.93$ for the low- and high-temperature data, respectively. Also the fit quality is improved, as now $\chi^2 = 22.27$  for  the low temperature data and $\chi^2 = 12.54$  for  the high temperature data. The relative improvement is bigger for the low temperature data (see Table\,\ref{SBN61-summary-15K-290K}, column 2), but again the quality of the fit is better for the high temperature data set.  Motivated by this success we included an additional modulation of the ADP's at the Sr2 and Ba sites, as this proved valuable in the single-crystal refinement  \cite{Woike-SBN61-2003}.  However, as can be seen from Table\,\ref{SBN61-summary-15K-290K} (column 3), the improvement over model II is marginal, and is not significant. 
We thus concentrate on the results of model II for the comparison of the two structures at 15\,K and 290\,K. 

\begin{figure}
 \includegraphics[width=7cm]{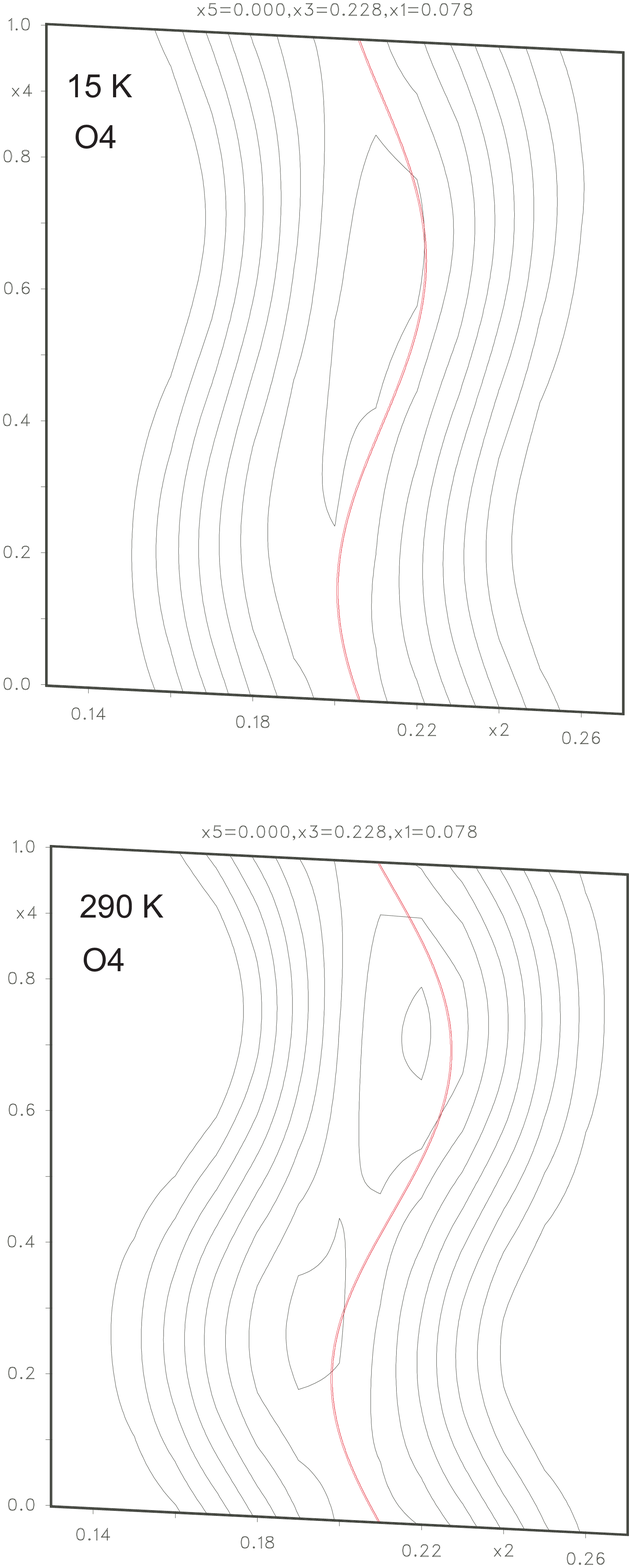}%
 \center
\caption{Color online) Modulation of O4 in SBN61 in the  x4/x2 plane at $T = 15$\,K (top) and  $T = 290$\,K (bottom). The O4 atom is marked by a bold (red) line.
The cut is through five-dimensional space as defined by Eqs. \ref{eq:def-rec}-\ref{eq:def-e2}, where x$_1$-x$_3$ is the standard reciprocal space.}
\label{SBN61-O4-modulation}
\end{figure}

Before comparing the differences in the modulation parameters at T = 15\,K and  T = 290\,K we like to stress again, that the description of the powder data with a modulated structure is feasible and necessary, although no separate satellite reflections could be measured. This is illustrated in Fig.\,\ref{SBN61-O4-modulation}, which shows observed Fourier maps for the atom O4 and the refined positional modulation as a function of the coordinate x4 corresponding to the first modulation vector q$_1$. The atom O4 is lying in the plane of the Sr and Ba atoms, which occupy the A1 and A2 channels in the unfilled tungsten bronze structure (see Fig. 1). Due to the incomplete filling and disorder of Sr/Ba in these channels, the adjacent oxygen-octahedra are tilted, leading to the observed incommensurate modulation \cite{Woike-SBN61-2003, schaniel-thesis-2003}. For the O4 atoms this results mainly in a positional modulation in the a-b-plane (coordinates x,y). Fig.\,\ref{SBN61-O4-modulation} illustrates this behaviour and shows that the refined modulation accounts satisfactorily for the observed modulation. Selected modulation parameters are given in Table\,\ref{SBN61-compare-modulation}. We observe small differences in the modulation between the 290\,K and 15\,K measurement, e.g., the modulation of the O4 atom (Fig.\,\ref{SBN61-O4-modulation}) is slightly larger in the y-direction at 290\,K. Similar small differences are observed for other directions and atoms. Notably here the displacement parameters of the O(4) and O(5) atom even increase with decreasing temperature. Furthermore we find that the occupations of the Sr(1) and Sr(2) position change slightly compared to the average structure, as the Sr(1) occupation increases to 0.695(9) and that of Sr(2) decreases to 0.415(4), and is hence in very good agreement with the single crystal results \cite{Woike-SBN61-2003}. The complete listing of refined parameters for the 15\,K and 290\,K data is given in Tables V-VIII in the supplementary material. Thereby s(1,0) and c(1,0) denote the sine and cosine part of the harmonic wave for the modulation vector \textbf{q$_1$} and s(0,1) and c(0,1) those of \textbf{q$_2$}. 

\begin{table}
\caption{Comparison of the positional modulation factors  (model II, Table. \ref{SBN61-summary-15K-290K}) for  Sr$_{0.61}$Ba$_{0.39}$Nb$_{2}$O$_{6}$, SBN61, measured on HRPT/SINQ at $\lambda = 1.886$\,\AA\enspace in the high-resolution mode as tabulated in Tabs. V-VI for 15K and Tabs. VII-VII for 290K in the supplementary material \cite{Schefer-Supplementary-Material-2006}.}
\label{SBN61-compare-modulation}
{\smallskip}
\begin{tabular}{ll|l|l}
\hline
 Atom     & parameter      &  15K                        & 290 K                   \\
\hline
O2        & z (s,1,0)      &  -0.0219(16)                 &  -0.0148(17)      \\
O2        & z (c,1,0)      &  -0.0156(18)                 &  -0.0100(17)       \\
O2        & z (s,0,1)      &  -0.0116(18)                 &  -0.0057(16)       \\
O2        & z (c,0,1)      &   0.0354(12)                 &  0.0211(14)       \\
\hline
O3        & z (s,1,0)      &  0.0161(16)                 & 0.0297(15)       \\
O3        & z (c,1,0)      &  0.0146(15)                 & 0.0054(16)      \\
O3        & z (s,0,1)      & 0.0174(14)                 & 0.0188(16)       \\
O3        & z (c,0,1)      &  -0.0158(15)                 &  -0.0171(16)       \\
\hline
O4        & x (s,1,0)      & 0.0037(14)                 & 0.0041(12)       \\
O4        & x (c,1,0)      & 0.0243(9)                 &  0.0251(8)       \\
O4        & x (s,0,1)      & 0.0135(13)                 & 0.0139(13)      \\
O4        & x (c,0,1)      & -0.0217(10)                 &  -0.0188(10)       \\
O4        & y (s,1,0)      & -0.0093(10)                 & -0.0142(9)       \\
O4        & y (c,1,0)      & -0.0055(9)                 &  -0.0033(9)       \\
O4        & y (s,0,1)      & -0.0081(10)                 & -0.0064(10)      \\
O4        & y (c,0,1)      & 0.0085(9)                 & 0.0094(9)       \\
\hline
O5        & x (s,1,0)      & 0.0117(12)                 & 0.0085(13)       \\
O5        & x (s,0,1)      & -0.0096(14)                 & -0.0056(15)      \\
\hline
O4        & U$_{iso}$      & 0.0104(15)                 & 0.0078(14)      \\
O5        & U$_{iso}$      & 0.038(3)                 & 0.0033(3)      \\
\hline
\end{tabular}
\end{table}

\section{Discussion}

The presented high-resolution diffraction data clearly show, that there is no change in the average structure of SBN61 in the temperature range 300-15\,K, especially no indications for a transition into a monoclinic or orthorhombic space group, as suggested in Ref. \cite{Xu-PRB-1989}, could be detected. The structural data are supported by the specific heat measurements, where neither a latent heat nor a significant change in the slope of $C_{p}$ vs T curve, as found in Ref. \cite{Xu-PRB-1989}, could be detected. The lattice parameter a shows the expected temperature dependence, as it increases steadily with increasing temperature. The temperature dependence of c reflects the relaxor behavior of the ferroelectric phase transition at $T_c \simeq 350$\,K. Notably these structural changes start already at  T = 200\,K and  end at about 400\,K, where the parameter c increases again up to 500\,K. These results are in excellent agreement with the x-ray diffraction results of Prokert \textit{et.al} \cite{prokert-SBN39-1995}, where also the decrease in c between 200\,K and 450\,K was observed. Hence the anisotropic thermal expansion reported by Qadri \textit{et al.} \cite{Qadri-APL-2005} in the temperature range 160-300\,K in SBN75 can be explained completely with the onset of the relaxor phase transition at $T_c \simeq 300 $\,K, bearing in mind that in SBN75 the phase transition temperature ($T_c \simeq 300 $\,K) is lower than in SBN61 ($T_c \simeq 350 $\,K) \cite{David-PSS-2004} and in complete agreement with the earlier results on SBN75 of Prokert \textit{et al.} \cite{prokert-SBN75-1998}

More insight in the origin of the transverse dielectric anomalies observed below 100\,K \cite{Ko-JAP-2002} might be obtained from the inspection of the modulated refinement on the two high-resolution powder patterns. The structural results obtained from our powder data at 290\,K are in reasonable 
agreement with the single crystal data \cite{schaniel-thesis-2003} and show, that the average structures
are not sufficient to study the SBN system (compare Table \ref {SBN61-summary-15K-290K} and Figs. \ref{SBN61-jana2000-15K} and \ref{SBN61-jana2000-15K-averaged}). As illustrated for the atom O4 the positional modulation is smaller at 15K compared to 290\,K (see Fig. \ref{SBN61-O4-modulation})  indicating that a small change in the amplitude of the incommensurate structure occurs upon cooling. Since the modulation is mainly due to the tilting of the NbO$_6$ octahedra, we would expect a change in their configuration or the freezing of a possible dynamic behavior such as a temperature driven rotation of the NbO$_6$ octahedra. This interpretation is in line with the argumentation of Ko \textit{et al.} \cite{Ko-JAP-2002}, who observed a temperature driven process with an activation energy of about 0.087\,eV  and an attempt frequency of $1.5\cdot10^{11}$\,Hz (Arrhenius like). Effectively a distribution of relaxation times was observed leading to a distribution of activation energies in the range 0.08-0.22\,eV, which might be explained by random environments of the oxygen octahedra due to the incommensurate structure. This explanation might also hold for the interpretation of the Raman data, as the authors of the Raman study stated that a change in the local symmetry of the NbO$_6$ octahedra would also explain their observations \cite{Faria-JRS-2003}. 
Ko and Kojima performed also low-temperature Brillouin spectroscopy on SBN61 and found that the coupling between the acoustic modes and the thermal activation process is very weak \cite{Ko-2002}. The observed mild softening in the temperature dependence of the frequency shifts of the acoustic modes was ascribed to a dynamical behavior rather than a structural phase transition. More detailed information about a possible phonon softening at low temperatures might be obtained from inelastic neutron measurements, which to our knowledge have not been performed so far.
The specific heat data can also be interpreted in terms of freezing in of some dynamics. 
The fact, that the specific heat of SBN61 can not be explained by a simple Debye $\beta T^3$ dependence was already observed in 1982 by Henning and co-workers \cite{henning-pss-1982}, which explained their findings in the temperature range 0.3-30\,K with a glass-like behavior. Note that this glassy behavior in the specific heat was used by De Yoreo \textit{et al.} to separate ferroelectrics in two classes \cite{DeYoreo-PRB-1985}.  Our experiments are in good agreement with these observations, as also we have to explain the low temperature part using three Einstein modes in addition to the Debye part. The glass-like features might well be due to the freezing of a rotational dynamics of the oxygen octahedra, which then solidifies in more or less arbitrary directions at low temperatures. Such a freezing could also explain the fact that the obtained fit quality of the incommensurate modulation is worse at low temperature, since a rotational dynamics at high temperature leads only to the smearing out of displacement parameters, while a frozen configuration at low temperature might considerably disturb the periodicity of the modulation. The harmonic approximation used for the description of the incommensurate modulation cannot account for such a distortion. However, an indication for an oxygen octahedra distortion are the rather large displacement parameters for the atoms O4 and O5 at 15\,K. A further hint to dynamical behavior is the intrinsic broadness of the satellite peaks as observed in single crystal neutron diffraction at room temperature \cite{Schaniel-Schefer-SBN61-superspace-2002,schaniel-thesis-2003}. However, the reason for this broadening is unknown at the moment, and a temperature dependent investigation has to show whether it vanishes at low temperatures, as indicated by the lower quality in the fit
shown in Table \ref{SBN61-summary-15K-290K}.

\section{Conclusions}

Neutron powder diffraction experiments show no structural phase transition of the average structure of SBN61. They give evidence that the dielectric anomalies observed below T = 100\,K are due to small changes in the amplitude of the incommensurate modulation of the oxygen octahedra, and hence support the interpretation of Ko \textit{et al.}\cite{Ko-JAP-2002}, that some barrier crossing processes such as the concerted rotation of the oxygen octahedra are responsible for the observed dielectric phenomena. This interpretation is supported by specific heat measurements, which exhibit glass-like features at low temperatures. Furthermore we showed that the structural changes connected with the high temperature ferroelectric phase transition ($T_c \simeq 350 $\,K) start already at about 200\,K and are responsible for the anisotropic thermal expansion in SBN crystals \cite{Qadri-APL-2005} in this temperature range.  \\

\section{Acknowledgments}
 Neutron beam  time at the high-resolution powder diffractometers HRPT and POLDI at the Swiss Spallation Neutron  Source SINQ, is gratefully acknowledged as well as the support by LDM group of PSI and the crystal growth department of the University of Osnabr\"uck. The development of the JANA2000 program package was supported by the Grant Agency of the Czeck Republic, grant 202/06/0757.\\


\clearpage


\begin{table*}
\caption{Comparison of the averaged structures of Sr$_{0.61}$Ba$_{0.39}$Nb$_{2}$O$_{6}$(model I) at 15\,K and 290\,K.}
\label{SBN61-compare-average}
 {\smallskip}
\begin{tabular}{lllllllll}
T (K) & Atom & occ.   & x& y & z& U$_{iso}$ & \\
\hline
15 & Nb1&1& 0& 0.5&-0.0211(18)& 0.0003(17)\\
290 & Nb1&1& 0& 0.5&-0.0179(18)& 0.0053(18)\\
15 & Nb2&1& 0.0737(2)& 0.2129(2)&-0.0194(13)& 0.0035(10)\\
290 & Nb2&1& 0.0736(2)& 0.2127(2)&-0.0176(14)& 0.0056(10)\\
15 & Sr1&0.656(14)& 0& 0& 0.229(3)& 0.006(3)\\
290 & Sr1&0.649(12)& 0& 0& 0.223(3)& 0.005(3)\\
15 & Ba@Sr2&0.4875& 0.1682(3)& 0.6682(3)& 0.2383& 0.026(2)\\
290 & Ba@Sr2&0.4875& 0.1693(3)& 0.6693(3)& 0.2383& 0.031(2)\\
15 & Sr2&0.434(7)& 0.1682& 0.6682& 0.2383& 0.026(2)\\
290 & Sr2&0.438(6)& 0.1693& 0.6693& 0.2383& 0.031(2)\\
15 & O1&1& 0.2174(3)& 0.2826(3)&-0.0468(18)& 0.0111(17)\\
290 & O1&1& 0.2169(3)& 0.2831(3)&-0.0427(19)& 0.0139(16)\\
15 & O2&1& 0.1386(3)& 0.0679(4)&-0.0475(15)& 0.0161(11)\\
290 & O2&1& 0.1389(3)& 0.0670(3)&-0.0422(15)& 0.0190(10)\\
15 & O3&1&-0.0066(2)& 0.3449(3)&-0.0416(13)& 0.0098(11)\\
290 & O3&1&-0.0067(2)& 0.3443(3)&-0.0394(14)& 0.0132(11)\\
15 & O4&1& 0.0830(4)& 0.2017(3)& 0.217(2)& 0.0370(15)\\
290 & O4&1& 0.0814(3)& 0.2031(3)& 0.221(2)& 0.0370(13)\\
15 & O5&1& 0& 0.5& 0.235(3)& 0.048(4)\\
290 & O5&1& 0& 0.5& 0.237(3)& 0.044(3) \\
 \hline
\end{tabular}
\end{table*}

\end{document}



\title{
Structural properties  in Sr$_{0.61}$Ba$_{0.39}$Nb$_{2}$O$_{6}$  in the temperature range 10 K to 500 K investigated by high-resolution neutron powder diffraction and specific heat measurements}

\author{J. Schefer$^1$, D. Schaniel$^2$, V. Pomjakushin$^1$, U. Stuhr$^1$, V. Pet\v{r}\'{\i}\v{c}ek$^3$, Th. Woike$^2$, M. W\"ohlecke$^4$ and  M. Imlau$^4$}
\email{jurg.schefer@psi.ch}
\affiliation{$^1$Laboratory for Neutron Scattering, ETHZ \& PSI, CH-5232 Villigen, PSI, Switzerland\\
$^2$Institut f\"ur Mineralogie, University at Cologne, D-50674 K\"oln, Germany\\
$^3$Institute of Physics, Academy of Sciences of the Czech Republic, Na Slovance 2,18221 Praha 8, Czech Republic\\
$^4$Fachbereich Physik, University of Osnabr\"uck, D-49069 Osnabr\"uck, Germany}

\date{\today}

\pacs{61.12.Ld 33.15.Dj}




\begin{abstract}
Additonal Material: Full tables and full calculated and observed neutron powder patterns.
\end{abstract}
\pacs{61.12.Ld 33.15.Dj}
\maketitle


\setcounter{table}{4}
\setcounter{figure}{11}

\pagebreak[4]


\begin{table*}[ht]
\caption{Positional parameters of the incommensurate structure (model II) of  Sr$_{0.61}$Ba$_{0.39}$Nb$_{2}$O$_{6}$ at 15K. R-values are tabulated in Tab. II, column 2. Ba atoms at the Sr2 postions are indicated as Ba@Sr2.}
\label{SBN61-15K-parameters}
 {\smallskip}
\begin{tabular}{lllllllll}
Atom & occ. & wave  & x& y & z& U$_{iso}$ & \\
\hline
Nb1&1&& 0& 0.5&-0.0046(16)& 0.0007(16) \\
   &&s,1,0&-0.0004(14)&-0.0004(14)& 0&\\
   &&c,1,0& 0& 0& 0.005(2)&\\
   &&s,0,1&-0.0037(14)& 0.0037(14)& 0&\\
   &&c,0,1& 0& 0& 0.016(2)&\\
Nb2&1&& 0.07465(16)& 0.21227(15)&-0.0093(12)& 0.0007(9)\\
   &&s,1,0& 0.0030(10)&-0.0023(8)& 0.0067(13)&\\
   &&c,1,0& 0.0034(11)&-0.0001(9)&-0.0002(10)&\\
   &&s,0,1& 0.0006(9)& 0.0018(8)&-0.0010(12)&\\
   &&c,0,1&-0.0006(11)&-0.0007(10)& 0.0007(13)&\\
Sr1&0.695(9)&& 0& 0& 0.224(2)&-0.004(2)\\
   &&s,1,0& 0.0040(12)&-0.0004(13)& 0&\\
   &&c,1,0& 0& 0& 0.004(4)&\\
   &&s,0,1&-0.0004(13)&-0.0040(12)& 0&\\
   &&c,0,1& 0& 0& 0.004(4)&\\
Ba@Sr2&0.4875&& 0.1710(3)& 0.6710(3)& 0.2383& 0.0132(18)\\
     &&s,1,0&-0.0041(9)&-0.0041(9)& 0.015(3)&\\
     &&c,1,0&-0.0033(11)&-0.0033(11)& 0.009(3)&\\
     &&s,0,1&-0.0059(10)& 0.0059(10)& 0&\\
     &&c,0,1&-0.0022(10)&-0.0022(10)&-0.006(3)&\\
Sr2&0.415(4)&& 0.171& 0.671& 0.2383& 0.0132(18)\\
   &&s,1,0&-0.0041(9)&-0.0041(9)& 0.015(3)&\\
   &&c,1,0&-0.0033(11)&-0.0033(11)& 0.009(3)&\\
   &&s,0,1&-0.0059(10)& 0.0059(10)& 0&\\
   &&c,0,1&-0.0022(10)&-0.0022(10)&-0.006(3)&\\
O1&1&& 0.2181(2)& 0.2819(2)&-0.0325(16)& 0.0007(18)\\
  &&s,1,0& 0.0015(13)& 0.0015(13)& 0&\\
  &&c,1,0&-0.0079(10)& 0.0079(10)&-0.0100(18)&\\
  &&s,0,1& 0.0063(12)&-0.0063(12)&-0.002(2)&\\
  &&c,0,1& 0.0030(12)&-0.0030(12)& 0.0101(18)&\\
O2&1&& 0.1398(3)& 0.0679(3)&-0.0309(15)& 0.0042(12)\\
  &&s,1,0& 0.0105(11)& 0.0028(10)&-0.0219(16)&\\
  &&c,1,0& 0.0042(15)& 0.0015(13)&-0.0156(18)&\\
  &&s,0,1& 0.0020(15)&-0.0016(13)&-0.0116(18)&\\
  &&c,0,1&-0.0071(12)&-0.0015(13)& 0.0354(12)&\\
O3&1&&-0.00725(19)& 0.3450(2)&-0.0328(14)& 0.0075(9)\\
  &&s,1,0& 0.0005(13)&-0.0014(12)& 0.0161(16)&\\
  &&c,1,0& 0.0000(14)& 0.0006(13)& 0.0146(15)&\\
  &&s,0,1&-0.0005(13)&-0.0011(13)& 0.0174(14)&\\
  &&c,0,1&-0.0028(12)& 0.0017(12)&-0.0158(15)&\\
O4&1&& 0.0791(4)& 0.2030(2)& 0.2285(16)& 0.0104(15)\\
  &&s,1,0& 0.0037(14)&-0.0093(10)& 0.004(3)&\\
  &&c,1,0& 0.0243(9)&-0.0055(9)&-0.001(3)&\\
  &&s,0,1& 0.0135(13)&-0.0081(10)&-0.002(3)&\\
  &&c,0,1&-0.0217(10)& 0.0085(9)&-0.008(2)&\\
O5&1&& 0& 0.5& 0.230(3)& 0.038(3)\\
  &&s,1,0& 0.0117(12)& 0.0117(12)& 0&\\
  &&c,1,0& 0& 0&-0.004(7)&\\
  &&s,0,1&-0.0096(14)& 0.0096(14)& 0&\\
  &&c,0,1& 0& 0& 0.003(7)&\\
\hline
\end{tabular}
\end{table*}

\begin{table*}[ht]
\caption{Temperature factors of the incommensurate structure of Sr$_{0.61}$Ba$_{0.39}$Nb$_{2}$O$_{6}$ at 15K. R-values are tabulated in Tab. II, column 2  (model II)} 
\label{SBN61-15K-temp}
 {\smallskip}
\begin{tabular}{lllllllll}
Atom & wave & U11& U22  & U33 & U12 & U13 & U23 \\
\hline
Nb1&& 0.0077(18)& 0.0077(18)&-0.013(4)&-0.0027(19)& 0& 0\\
Nb2&& 0.0113(18)&-0.0059(12)&-0.0032(15)& 0.0020(12)& 0.005(2)&-0.004(2)\\
Sr1&&-0.015(2)&-0.015(2)& 0.018(6)& 0& 0& 0\\
Ba@Sr2&& 0.018(2)& 0.018(2)& 0.004(4)&-0.008(2)&-0.021(3)&-0.021(3)\\
Sr2&& 0.018(2)& 0.018(2)& 0.004(4)&-0.008(2)&-0.021(3)&-0.021(3)\\
\hline
\end{tabular}
\end{table*}

\pagebreak[4]

\begin{figure} \includegraphics[width=23cm,angle=90]{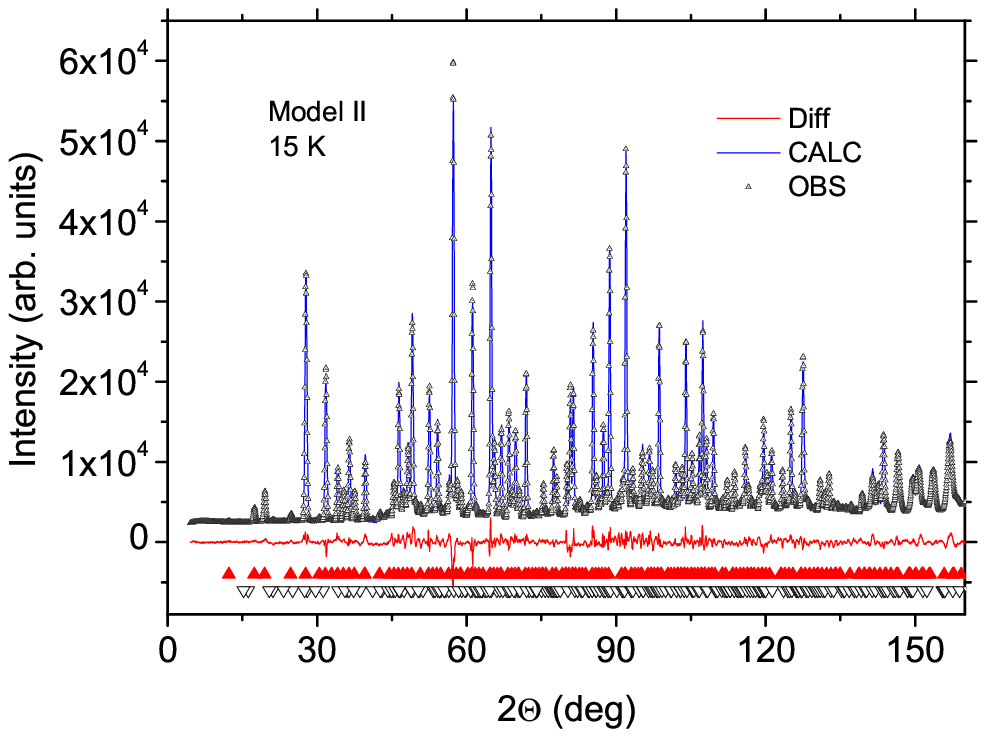}%
\caption{Graphical result of the Rietveld \cite {petricek2000} refinement of SBN61 powder data collected on HRPT using the high-resolution mode, $\lambda = 1.886$\,\AA, at temperature T=290K. Detailed  results are listed in Tabs.
 \ref{SBN61-15K-parameters},\ref{SBN61-15K-temp} (incommensurate model I).
 HKL's are marked by closed triangles(average structure: top row) and by open triangles (incommensurate satellites, lower row).
 }
\label{SBN61-jana2000-15K}
\end{figure}

\newpage
\pagebreak[4]


\begin{table*}[ht]
\caption{Positional parameters of the incommensurate structure of Sr$_{0.61}$Ba$_{0.39}$Nb$_{2}$O$_{6}$ at 290K. R-values are tabulated in Tab. II, column 5 (model II).}
\label{SBN61-290K-parameters}
 {\smallskip}
\begin{tabular}{lllllllll}
Atom & occ. & wave & x& y & z&  U$_{iso}$ & \\
\hline
Nb1&1&& 0& 0.5&-0.013(2)& 0.0038(17)\\
   &&s,1,0& 0.0015(16)& 0.0015(16)& 0&\\
   &&c,1,0& 0& 0& 0.003(2)&\\
   &&s,0,1& 0.0052(17)&-0.0052(17)& 0&\\
   &&c,0,1& 0& 0& 0.010(2)&\\
Nb2&1&& 0.07427(18)& 0.21243(16)&-0.0166(15)& 0.0081(10)\\
   &&s,1,0&-0.0008(11)& 0.0002(11)& 0.0017(15)&\\
   &&c,1,0& 0.0016(13)& 0.0047(13)& 0.0016(13)&\\
   &&s,0,1& 0.0016(11)&-0.0004(11)&-0.0002(13)&\\
   &&c,0,1& 0.0012(11)& 0.0022(10)& 0.0020(12)&\\
Sr1&0.695(9)&& 0& 0& 0.228(3)&-0.002(3)\\
   &&s,1,0& 0.0001(11)&-0.0020(12)& 0&\\
   &&c,1,0& 0& 0&-0.017(4)&\\
   &&s,0,1&-0.0020(12)&-0.0001(11)& 0&\\
   &&c,0,1& 0& 0&-0.017(4)&\\
Ba@Sr2&0.4875&& 0.1718(3)& 0.6718(3)& 0.2383& 0.0176(15)\\
     &&s,1,0&-0.0012(9)&-0.0012(9)& 0.002(3)&\\
     &&c,1,0&-0.0036(9)&-0.0036(9)&-0.006(2)&\\
     &&s,0,1&-0.0035(12)& 0.0035(12)& 0&\\
     &&c,0,1& 0.0036(10)& 0.0036(10)&-0.007(3)&\\
Sr2&0.415(4)&& 0.1718& 0.6718& 0.2383& 0.0176(15)\\
   &&s,1,0&-0.0012(9)&-0.0012(9)& 0.002(3)&\\
   &&c,1,0&-0.0036(9)&-0.0036(9)&-0.006(2)&\\
   &&s,0,1&-0.0035(12)& 0.0035(12)& 0&\\
   &&c,0,1& 0.0036(10)& 0.0036(10)&-0.007(3)&\\
O1&1&& 0.2181(2)& 0.2819(2)&-0.025(2)& 0.0099(17)\\
  &&s,1,0& 0.0015(16)& 0.0015(16)& 0&\\
  &&c,1,0& 0.0058(15)&-0.0058(15)&-0.007(2)&\\
  &&s,0,1&-0.0026(17)& 0.0026(17)&-0.005(2)&\\
  &&c,0,1&-0.0042(14)& 0.0042(14)& 0.0093(19)&\\
O2&1&& 0.1396(3)& 0.0675(3)&-0.0401(15)& 0.0137(9)\\
  &&s,1,0& 0.0027(15)& 0.0004(15)&-0.0148(17)&\\
  &&c,1,0& 0.0026(15)&-0.0002(15)&-0.0100(17)&\\
  &&s,0,1&-0.0010(18)&-0.0006(15)&-0.0057(16)&\\
  &&c,0,1& 0.0030(13)& 0.0007(14)& 0.0211(14)&\\
O3&1&&-0.00704(19)& 0.3450(3)&-0.0293(18)& 0.0088(11)\\
  &&s,1,0& 0.0018(14)&-0.0029(14)& 0.0297(15)&\\
  &&c,1,0&-0.0014(16)&-0.0007(17)& 0.0054(16)&\\
  &&s,0,1& 0.0044(16)&-0.0004(14)& 0.0188(16)&\\
  &&c,0,1& 0.0018(17)&-0.0006(18)&-0.0171(16)&\\
O4&1&& 0.0788(4)& 0.2035(2)& 0.232(2)& 0.0078(14)\\
  &&s,1,0& 0.0041(12)&-0.0142(9)&-0.001(3)&\\
  &&c,1,0& 0.0251(8)&-0.0033(9)& 0.000(3)&\\
  &&s,0,1& 0.0139(13)&-0.0064(10)& 0.002(3)&\\
  &&c,0,1&-0.0188(10)& 0.0094(9)& 0.004(2)&\\
O5&1&& 0& 0.5& 0.234(3)& 0.033(3)\\
  &&s,1,0& 0.0085(13)& 0.0085(13)& 0&\\
  &&c,1,0& 0& 0& 0.002(6)&\\
  &&s,0,1&-0.0056(15)& 0.0056(15)& 0&\\
  &&c,0,1& 0& 0& 0.007(6)&\\
\hline
\end{tabular}
\end{table*}

\begin{table*}[ht]
\caption{Temperature factors of the incommensurate structure of  Sr$_{0.61}$Ba$_{0.39}$Nb$_{2}$O$_{6}$ at 290K. R-values are tabulated in Tab. II, column 5 (model II).}  
\label{SBN61-290K-temp}
 {\smallskip}
\begin{tabular}{lllllllll}
\hline
Atom & wave & U11& U22  & U33 & U12 & U13 & U23 \\
\hline
Nb1&& 0.008(2)& 0.008(2)&-0.005(4)&-0.004(2)& 0& 0\\
Nb2&& 0.0188(18)&-0.0037(15)& 0.0093(19)& 0.0000(15)& 0.031(2)&-0.004(3)\\
Sr1&&-0.006(3)&-0.006(3)& 0.005(10)& 0& 0& 0\\
Ba@Sr2&& 0.030(2)& 0.030(2)&-0.007(3)&-0.018(2)&-0.005(4)&-0.005(4)\\
Sr2&& 0.030(2)& 0.030(2)&-0.007(3)&-0.018(2)&-0.005(4)&-0.005(4)\\
\hline
\end{tabular}
\end{table*}

\pagebreak[4]

\begin{figure} [ht]\includegraphics[width=23cm,angle=90]{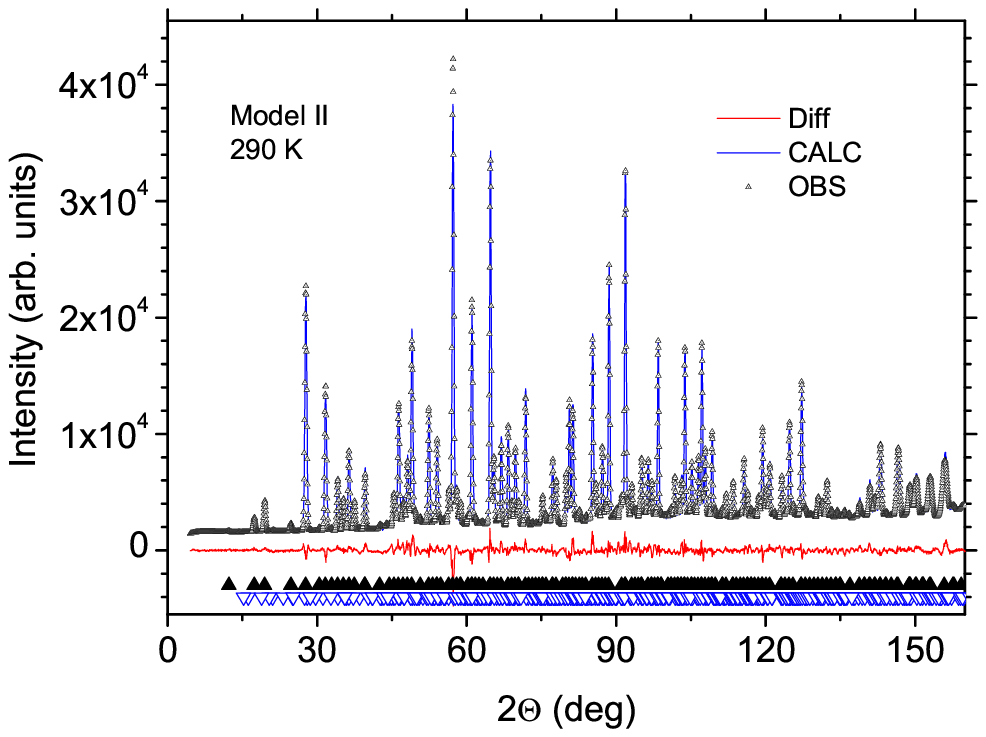}%
\caption{Graphical result of incommensurate structure (model II) using  Rietveld \cite {petricek2000} refinement. SBN61 powder data collected on HRPT using the high-resolution mode, $\lambda = 1.886$\,\AA, at temperature T=290K. Detailed  results are listed in Tabs. \ref{SBN61-290K-parameters},\ref{SBN61-290K-temp}
(incommensurate model I).
HKL's are marked by closed triangles(average structure: top row) and by open triangles (incommensurate satellites, lower row).}
\label{SBN61-jana2000-290K} 
\end{figure}

\newpage
\pagebreak[4]


\begin{table*}[ht]
\caption{Positional parameters of the averaged structure of Sr$_{0.61}$Ba$_{0.39}$Nb$_{2}$O$_{6}$ at 15K. R-values are tabulated in Tab. II, column 5 (model I).}
\label{SBN61-15K-average-parameters}
 {\smallskip}
\begin{tabular}{lllllllll}
Atom & occ. & wave & x& y & z&  U$_{iso}$ & \\
\hline
Nb1&1&-& 0& 0.5&-0.0211(17)& 0.0003(17)\\
Nb2&1&-& 0.0737(2)& 0.2129(2)&-0.0194(13)& 0.0035(10)\\
Sr1&0.656(14)&-& 0& 0& 0.229(3)& 0.006(3)\\
Ba@Sr2&0.4875&-& 0.1682(3)& 0.6682(3)& 0.2383& 0.026(2)\\
Sr2&0.434(7)&-& 0.1682& 0.6682& 0.2383& 0.026(2)\\
O1&1&-& 0.2174(3)& 0.2826(3)&-0.0468(18)& 0.0111(17)\\
O2&1&-& 0.1386(3)& 0.0679(4)&-0.0475(15)& 0.0161(11)\\
O3&1&-&-0.0066(2)& 0.3449(3)&-0.0416(13)& 0.0098(11)\\
O4&1&-& 0.0830(4)& 0.2017(3)& 0.217(2)& 0.0370(15)\\
O5&1&-& 0& 0.5& 0.235(3)& 0.048(4)\\
\hline
\end{tabular}
\end{table*}

\begin{table*}[ht]
\caption{Temperature factors of the averaged structure of  Sr$_{0.61}$Ba$_{0.39}$Nb$_{2}$O$_{6}$ at 15K. R-values are tabulated in Tab. II, column 5 (model I).}  
\label{SBN61-15K-average-temp}
 {\smallskip}
\begin{tabular}{lllllllll}
\hline
Atom & wave & U11& U22  & U33 & U12 & U13 & U23 \\
\hline

Nb1&-&-0.0008(19)&-0.0008(19)& 0.003(4)&-0.014(2)& 0& 0\\
Nb2&-& 0.0103(19)&-0.0003(16)& 0.001(2)& 0.0016(16)& 0.008(3)& 0.026(2)\\
Sr1&-&-0.007(3)&-0.007(3)& 0.031(8)& 0& 0& 0\\
Ba@Sr2&-& 0.026(3)& 0.026(3)& 0.024(6)&-0.018(3)&-0.032(3)&-0.032(3)\\
Sr2&-& 0.026(3)& 0.026(3)& 0.024(6)&-0.018(3)&-0.032(3)&-0.032(3)\\
\hline
\end{tabular}
\end{table*}

\pagebreak[4]

\begin{figure}[ht] \includegraphics[width=23cm,angle=90]{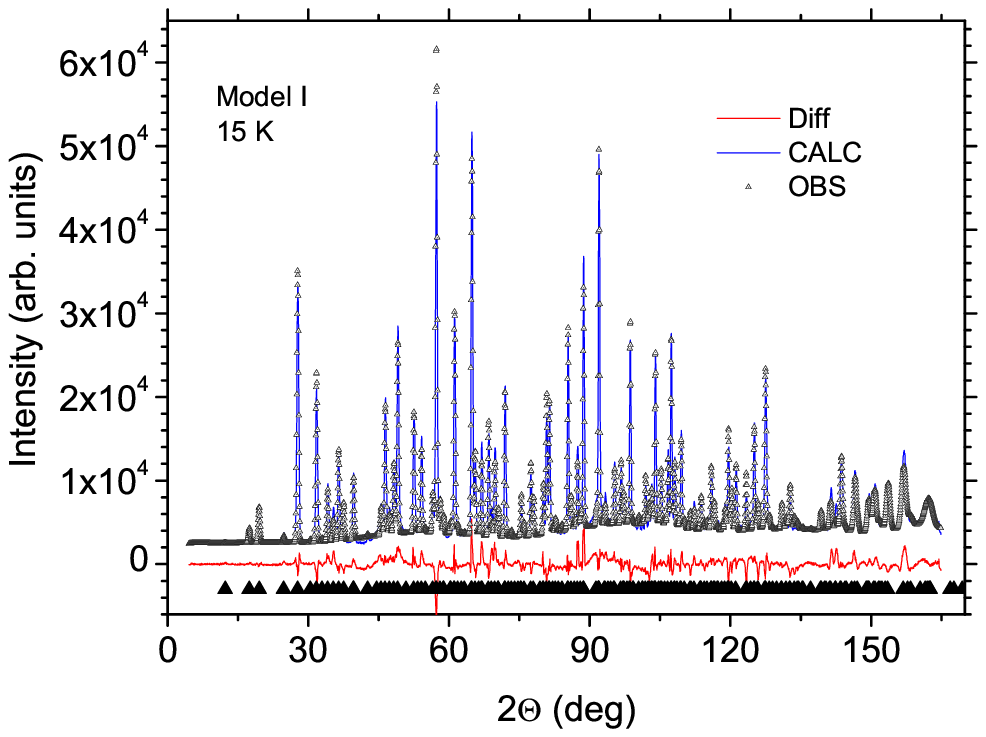}%
\caption{Graphical result of the averaged structure (model I) using  Rietveld  \cite {petricek2000} refinement of SBN61 powder data collected on HRPT using the high-resolution mode, $\lambda = 1.886$\,\AA, at temperature T=290K. Detailed  results are listed in Tabs. \ref{SBN61-15K-average-parameters},\ref{SBN61-15K-average-temp}.
HKL's are marked by vertical bars.}
\label{SBN61-jana2000-average-15K}
\end{figure}

\newpage
\pagebreak[4]


\begin{table*} [ht]
\caption{Positional parameters of the averaged structure of Sr$_{0.61}$Ba$_{0.39}$Nb$_{2}$O$_{6}$ at 290K. R-values are tabulated in Tab. II, column 4 (model I).}
\label{SBN61-290K-average-parameters}
 {\smallskip}
\begin{tabular}{lllllllll}
Atom & occ. & wave & x& y & z&  U$_{iso}$ & \\
\hline
Nb1&1&-& 0& 0.5&-0.0179(18)& 0.0053(18)\\
Nb2&1&-& 0.0736(2)& 0.2127(2)&-0.0176(14)& 0.0056(10)\\
Sr1&0.649(12)&-& 0& 0& 0.223(3)& 0.005(3)\\
Ba@Sr2&0.4875&-& 0.1693(3)& 0.6693(3)& 0.2383& 0.031(2)\\
Sr2&0.438(6)&-& 0.1693& 0.6693& 0.2383& 0.031(2)\\
O1&1&-& 0.2169(3)& 0.2831(3)&-0.0427(19)& 0.0139(16)\\
O2&1&-& 0.1389(3)& 0.0670(3)&-0.0422(15)& 0.0190(10)\\
O3&1&-&-0.0067(2)& 0.3443(3)&-0.0394(14)& 0.0132(11)\\
O4&1&-& 0.0814(3)& 0.2031(3)& 0.221(2)& 0.0370(13)\\
O5&1&-& 0& 0.5& 0.237(3)& 0.044(3)\\
\hline
\end{tabular}
\end{table*}

\begin{table*}[ht]
\caption{Temperature factors of the averaged structure of  Sr$_{0.61}$Ba$_{0.39}$Nb$_{2}$O$_{6}$ at 290K. R-values are tabulated in Tab. II, column 4 (model I).}  
\label{SBN61-290K-average-temp}
 {\smallskip}
\begin{tabular}{lllllllll}
\hline
Atom & wave & U11& U22  & U33 & U12 & U13 & U23 \\
\hline
Nb1&-& 0.005(2)& 0.005(2)& 0.007(5)&-0.012(2)& 0& 0\\
Nb2&-& 0.0112(18)& 0.0000(14)& 0.0054(19)&-0.0005(15)& 0.003(3)& 0.021(3)\\
Sr1&-&-0.004(3)&-0.004(3)& 0.024(7)& 0& 0& 0\\
Ba@Sr2&-& 0.036(3)& 0.036(3)& 0.020(5)&-0.021(3)&-0.026(4)&-0.026(4)\\
Sr2&-& 0.036(3)& 0.036(3)& 0.020(5)&-0.021(3)&-0.026(4)&-0.026(4)\\
\hline
\end{tabular}
\end{table*}

\pagebreak[4]

\begin{figure}[ht] \includegraphics[width=23cm,angle=90]{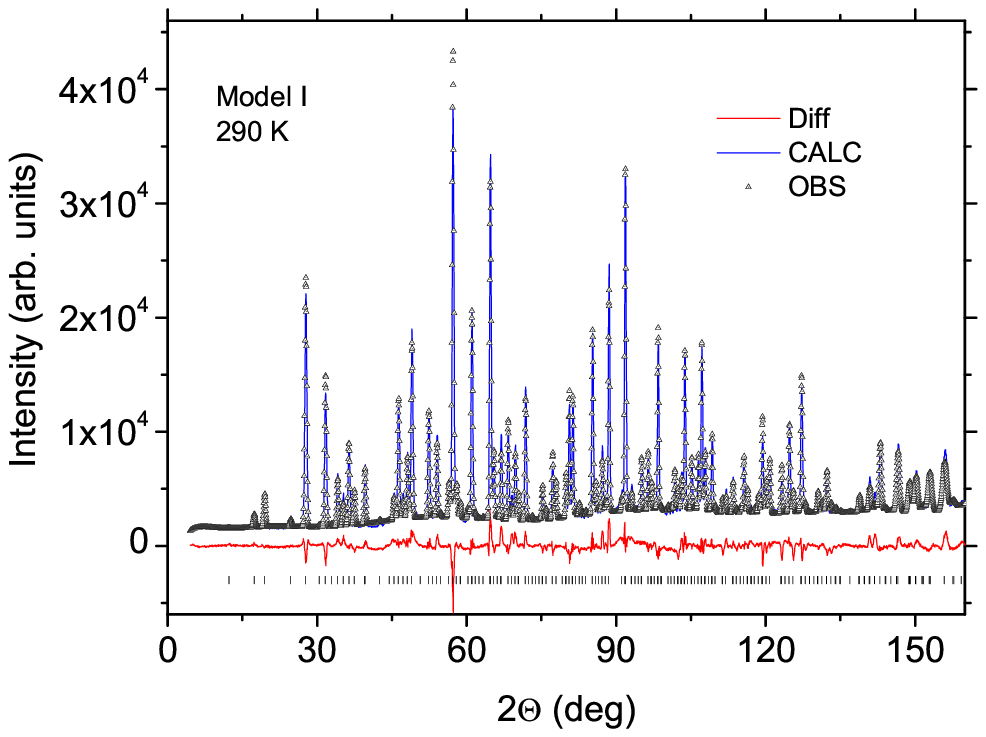}%
\caption{Graphical result of the averaged structure (model I) using  Rietveld  \cite {petricek2000} refinement of SBN61 powder data collected on HRPT using the high-resolution mode, $\lambda = 1.886$\,\AA, at temperature T=290K. Detailed  results are listed in Tabs. \ref{SBN61-290K-average-parameters},\ref{SBN61-290K-average-temp}.
HKL's are marked by vertical bars.}
\label{SBN61-jana2000-average-290K}
\end{figure}


\newpage
\pagebreak[4]

